\documentstyle[12pt,graphicx]{article}
\begin{document}
\title{ On meson masses}
\author{G.V.Efimov  \vspace*{0.2\baselineskip}\\
 \itshape Bogoliubov Laboratory of Theoretical Physics,\\
 \itshape Joint Institute for Nuclear Research,\\
{\it 141980 Dubna, Russia}\vspace*{0.2\baselineskip} }
%
%
\maketitle
\begin{abstract}
It is shown that in the framework of analytical confinement,
when quark and gluon propagators are induced by an vacuum selfdual
gluon field with constant strength, the masses of meson with quantum
numbers $Q=J^P$ and quark constituents $m_1,~m_2$ are described
with reasonable accuracy by the formula
$$ M_Q(m_1,m_2)=(m_1+m_2)\left[1+{A_Q\over
(m_1^2+1.13m_1m_2+m_2^2)^{0.625}}\right],$$
where a constant positive parameter $A_Q$ is unique for all mesons
with quantum numbers $Q=J^P$.

Sets of mesons $J^P=0^-,~1^-,~0^+,~1^+,~2^+,~3^-$ and different
flavors constituent quarks $(u=d,~s,~,c~,b)$ are considered.
\end{abstract}

\section{Introduction.}

Our basic point of view is that the confinement and hadronization
of quarks take place on the same distances. It means that the
knowledge of behavior of quark and gluon propagators on these
distances give us direct way to calculate meson masses using
the Bethe-Salpeter equation in ladder approximation if the
coupling constant $\alpha_s$ is small enough.

Within the standard approaches of QCD the quark and gluon propagators
in the confinement region are not obtained yet, so that different
symmetry arguments leading to some effective theory are used to
avoid direct calculations.

In the papers \cite{EfNed,EfKNed} we developed the approach based on
the assumption that the QCD vacuum is realized by the self-dual
homogeneous vacuum gluon field which is the classical solution of
the Yang-Mills equations. The propagators of quarks and gluons in
this field can be calculated in the explicit form and they are
entire analytic functions in $p^2$-complex plain. It means that
the self-dual homogeneous vacuum gluon field leads to the analytic
confinement of quarks. Manifestation of this gluon configuration in
the spectrum and weak decays of light mesons, their exited states,
glueballs were studied in papers \cite{EfBNed,EfB}. Our calculations
showed that the self-dual homogeneous gluon field can be considered
as a good candidate to realize QCD gluon vacuum.

However any calculations in this model are quite cumbersome so
that it is difficult to "see forest among trees".  I always wanted
to find some simple formulas which could have some general physical
consequences. I think this idea is realized in present paper.

Physical arguments are the following. Bound states are formed on
large distances where confinement plays the main role. So that
the correct description of confinement should lead to reasonable
description of bound states. Mesons as quark-antiquark bound states
are characterized by different quantum numbers which should be taken
into account for detailed description of meson characteristics.
However if the analytical confinement really takes place then one
can get some general dynamic properties of meson spectrum which
weakly depend on quantum numbers.

We use the functional integral techniques to calculate mass
spectrum of bound states. In this techniques representation of the
Bethe-Salpeter kernel with propagators, which are entire analytical
functions of the gaussian type, has some specific features leading
to consequences which can be checked on the real meson spectrum.
Namely this kernel is product of a polynomial depending on quantum
numbers and an exponent which is function of quark masses and mass
of the bound state under consideration and which gives the main
contribution into integral. Thus we can expect that in the lowest
semiquantative approximation the meson masses are defined by an
universal function depending on masses of constituent quarks and
dependence on quantum numbers can be approximated by a constant.

Let us consider the Particle Data Group \cite{PDG}. One can
extract some sets of mesons having the same quantum numbers $Q=J^P$,
but different flavors constitutes. There are five sets of mesons
having as constituents all four quarks $u=d,~s,~c,~b$. These sets
are: pseudoscalar $0^-$ - ten mesons, vector $1^-$ - eight mesons,
scalar $0^+$ - four mesons, axial $1^+$ - five mesons,
tensor $2^+$ - six mesons (see Section 3). In addition we
consider tensor multiplet $3^-$, containing three light mesons.

Thus we expect that the masses of these mesons with quantum numbers
$Q=J^P$ and quark constituents $m_1,~m_2$ are described by the formula
\begin{eqnarray}
\label{Muniv}
&& M_Q(m_1,m_2)={\cal M}(m_1,m_2,A_Q),
\end{eqnarray}
where ${\cal M}(m_1,m_2,A_Q)$ is an universal function, form of which
does not depend on quantum state. A constant parameter $A_Q$ defines
quantum state and it is unique for all mesons with quantum number
$Q=J^P$.

The aim of this paper is to show that the representation (\ref{Muniv})
really takes place and to clarify additional conditions providing this
formula.

\section{Bethe-Salpeter equation, \\
analytical confinement and meson masses}

The equation on the mass $M$ of a bound state of two constituent quarks
with masses $m_1$ and $m_2$ in framework of the Bethe-Salpeter equation
in ladder approximation looks as (see \cite{EfNed})
\begin{eqnarray}
\label{Mbound}
&&1={\alpha_s\over\pi}\int\!\!\!\int dy_1dy_2
V_Q(y_1)\Pi_Q(y_1-y_2|p)V_Q(y_2).
\end{eqnarray}
Here the polarization operator is
\begin{eqnarray}
\label{PQ}
&&\Pi_Q(y_1-y_2|p)\\
&&=\int dx~e^{i(px)}{\rm Tr}\left[
\Gamma_Q S_1(x+\mu_2(y_1-y_2))\Gamma_Q S_2(x-\mu_1(y_1-y_2))\right]
\nonumber
\end{eqnarray}
where the quark-antiquark vertex $\Gamma_Q$ defines the quantum numbers
of the bound state under consideration.

All calculations will be performed in the Euclidean space ${\bf R}^4$.

The quark propagator is chosen in the form induced by a vacuum
self-dual gluon field with constant strength (see details in \cite{EfNed})
\begin{eqnarray}
\label{Pqu}
S(y,m)&=&
{\Lambda^2\over8\pi^2}\int\limits_0^1 {du\over u^2}\tilde{S}(y,m,u)
\cdot e^{-{\Lambda^2y^2\over2u}}\cdot
\left({1-u\over 1+u}\right)^{{m^2\over4\Lambda^2}}~.
\end{eqnarray}
The parameter $\Lambda$ defines the confinement scale. The explicit form
of the polynomial $\tilde{S}(y,m,u)$ over variables $y$ and $m$ will be
not important in our calculations here but it can be found in \cite{EfNed}.

The other parameters are
\begin{eqnarray*}
&& p^2=-M^2,~~~~~\mu={M\over m_1+m_2},~~~~~\mu_1={m_1\over m_1+m_2},~~~~~
\mu_2={m_2\over m_1+m_2}.
\end{eqnarray*}

The vertex function $V_Q$ is determined by the solution of the
Bethe-Salpeter equation, but with acceptable accuracy it can be
approximated on large distances $y^2\sim{1\over\Lambda^2}$ by
\begin{eqnarray}
\label{VQ}
&&V_Q(y)\sim D(y) \sim V~e^{-{\Lambda^2\over2}y^2}.
\end{eqnarray}

The QCD coupling constant is defined by the standard way
\begin{eqnarray}
\label{cc}
&& \alpha_s={g^2\over4\pi}~.
\end{eqnarray}

Thus the mass of a bound state with quantum number $Q$ is defined by
the equation
\begin{eqnarray}
\label{MassEq}
1&=&{\alpha_s\over\pi}\int dy~V^2e^{-{y^2\over4}}
\int dx~e^{i(px)}{\rm Tr}\left[
\Gamma_Q S_1(x+\mu_2y)\Gamma_Q S_2(x-\mu_1y)\right]\\
&=&{\alpha_s\over\pi}\int\!\!\!\int\limits_0^1 du_1du_2
P_Q(u_1,u_2,\mu,\mu_1,\mu_2)e^{{(m_1+m_2)^2\over2\Lambda^2}
E(\mu,\mu_1,\mu_2,u_1,u_2)}.\nonumber
\end{eqnarray}
Here $P_Q(u_1,u_2,\mu,\mu_1,\mu_2)$ is a polynomial over parameters
$\mu,~\mu_1,~\mu_2$ and its explicit form is defined by the spin
structure of verteces and quark propagators. The quantum numbers
$Q=J^{PC}$ of bound states are defined by this polynomials.
The explicit form of this polynomials will be not important in our
arguments.

The function
\begin{eqnarray}
\label{E}
&&E(\mu,\mu_1,\mu_2,u_1,u_2)\\
&&=\mu^2\cdot{u_1u_2+2(\mu_1^2u_1+\mu_2^2u_2)
\over u_1+u_2+2}-
{\mu_1^2\over2}\ln\left({1+u_1\over1-u_1}\right)-
{\mu_2^2\over2}\ln\left({1+u_2\over1-u_2}\right).\nonumber
\end{eqnarray}
plays the main role in the equation (\ref{MassEq}) and
its behavior defines the meson mass spectrum. It is important that
this function depends on masses of constituent quarks $m_1$ and $m_2$
and does not depend on quantum numbers of bound states, so that we can
hope that the main features of a set of mesons with the same quantum
number $Q=J^{P}$ are defined by masses of constituent quarks  only.

Let us consider the behavior of the function $E$. Let us note bounds
of changing of our parameters:
\begin{eqnarray}
\label{param}
&& 0\leq\mu={M\over m_1+m_2}\leq1\div2.5;~~~~~0\leq\mu_j\leq1,~~~~
\mu_1+\mu_2=1.
\end{eqnarray}

In the case $\mu={M\over m_1+m_2}<1$ (the meson mass is less then sum
of masses of constituent quarks $m_1+m_2$), the function $E$ is
negative and main contribution into the integral (\ref{MassEq})
gives the vicinity of the point $u_1=0$, $u_2=0$. Then
$$e^{{(m_1+m_2)^2\over2\Lambda^2}E(\mu,\mu_1,\mu_2)}<1$$
and the solution of the mass equation requires strong coupling regime
$\alpha_s>1$.

In the case $\mu={M\over m_1+m_2}>1$ (the meson mass $M$ is more
then sum of masses of constituent quarks $m_1+m_2$), the function $E$
is positive and has a positive maximum in a point $0<u_1^{(0)}<1$,
$0<u_2^{(0)}<1$. Thus one can write approximately
\begin{eqnarray*}
&&\int\!\!\!\int\limits_0^1 du_1du_2
P_Q(u_1,u_2,\mu,\mu_1,\mu_2)e^{{(m_1+m_2)^2\over2\Lambda^2}
E(\mu,\mu_1,\mu_2,u_1,u_2)}\\
&&\approx C_Q e^{{(m_1+m_2)^2\over2\Lambda^2}{\cal E}(M,m_1,m_2)}
\end{eqnarray*}
where
\begin{eqnarray}
\label{Emax}
&& {\cal E}(M,m_1,m_2)=E(\mu,\mu_1,\mu_2)=\max\limits_{u_1,~u_2}
E(\mu,\mu_1,\mu_2,u_1,u_2).
\end{eqnarray}
The parameter $C_Q$ contains all information about quantum numbers of
states under consideration. We can suppose that $C_Q$ depends weakly
on mass parameters $\mu,~\mu_1,~\mu_2$.

Calculations of the maximum in (\ref{Emax}) in the region (\ref{param})
give the approximate formula
\begin{eqnarray}
\label{Eap}
&& (m_1+m_2)^2{\cal E}(M,m_1,m_2)\\
&&\approx\left[1.37m_1^2+1.55m_1m_2+1.37m_2^2\right]
\left({M\over m_1+m_2}-1\right)^{1.6}.\nonumber
\end{eqnarray}

Thus the equation (\ref{MassEq}) can be approximated by
\begin{eqnarray}
\label{MEqAp}
1&\sim&{\alpha_s\over\pi}C_Q
e^{{(m_1+m_2)^2\over2\Lambda^2}{\cal E}(M,m_1,m_2)}.
\end{eqnarray}

The solution of this mass equation requires weak coupling regime
$\alpha_s<1$. Finally the formula giving the mass of meson, which is
in a quantum state $Q=J^P$ and is constructing by quarks with masses
$m_1$ and $m_2$, is
\begin{eqnarray}
\label{MQap}
&& M_Q(m_1,m_2)\approx(m_1+m_2)\left[1+
{A_Q\over\left(m_1^2+1.13m_1m_2+m_2^2\right)^{0.625}}\right].
\end{eqnarray}
Here the positive constant $A_Q$ is the same for all mesons with
a given quantum number $Q=J^{P}$.

Now we should check this formula on real meson spectrum.

\section{Meson masses.}

Let us choose in the Table Particle Group Data \cite{PDG} mesons
with the same quantum numbers but with different flavors masses.
Mesons $J^P$ can be represented in the form
\begin{eqnarray}
\label{JP}
&& Q=J^{P}~\Rightarrow~
\left(\begin{array}{cccc}
uV_Q\bar{u} & uV_Q\bar{s} & uV_Q\bar{c} & uV_Q\bar{b}\\
 & sV_Q\bar{s} & sV_Q\bar{c} & sV_Q\bar{b}\\
 & & cV_Q\bar{c} & cV_Q\bar{b}\\
 & & & bV_Q\bar{b}\\
\end{array}\right).
\end{eqnarray}
We can choose only five sets of mesons with fixed $J^P$ having all
four constituent quarks $u=d,~s,~c,~b$
\begin{eqnarray*}
&& 0^-~\Rightarrow~P=
\left(\begin{array}{cccc}
\eta(547) & K(494) & D(1869) & B(5279)\\
 & \eta'(957) & D_s(1968) & B_s(5369)\\
 & & \eta_c(2979) & B_c(6400)\\
 & & & \eta_b(9300)\\
\end{array}\right);\\
&& 1^-~\Rightarrow~V=
\left(\begin{array}{cccc}
\omega(782) & K^*(892) & D^*(2007) & B^*(5325)\\
 & \phi(1020) & D_s^*(2112) & - \\
 & & J/\psi(3100) & - \\
 & & & \Upsilon(9460)\\
\end{array}\right);\\
&& 0^+~\Rightarrow~S=
\left(\begin{array}{cccc}
f_0(980) & - & - & - \\
 & f_0(1370) & - & - \\
 & & \chi_{c0}(3415) & - \\
 & & & \chi_{b0}(9893) \\
\end{array}\right);\\
&& 1^+~\Rightarrow~A=
\left(\begin{array}{cccc}
a_1(1260) & K_1(1270\div1400) & - & - \\
 & f_1(1420) & - & - \\
 & & \chi_{c1}(3510) & - \\
 & & & \chi_{b1}(9892)\\
\end{array}\right);\\
&& 2^+~\Rightarrow~D=
\left(\begin{array}{cccc}
f_2(1270) & K_2^*(1430) & D_2^*(2460) & - \\
 & f_2'(1525) & - & - \\
 & & \chi_{c2}(3556) & - \\
 & & & \chi_{b2}(9912)\\
\end{array}\right).
\end{eqnarray*}
In addition we consider three mesons $3^-$:
\begin{eqnarray*}
&& 3^-~\Rightarrow~T=
\left(\begin{array}{cc}
\omega_3(1670) & K_3^*(1780) \\
 & \phi_3(1850) \\
\end{array}\right).
\end{eqnarray*}

We want to show that the masses of all these 36 mesons with reasonable
accuracy are described the formula (\ref{MQap}). Our parameters which
should be determined by fitting are masses of constituent quarks
$m_u=m_d$, $m_s$, $m_c$, $m_b$ (4 parameters) and parameters
$A_P$, $A_V$, $A_S$, $A_A$, $A_D$, $A_T$ (6 parameters).

Our calculations consist of the following steps.

The first step. We calculate the minimum
\begin{eqnarray}
\label{Min1}
&& \Omega_Q=\min\limits_{m_f,~A_Q}
\sum\limits_{f:u,s,c,b}\left[1-
{M_{Q,m_f,m_f}(m_f,m_f,A_Q)\over (M_{Q,m_f,m_f})_{\rm exp}}\right]^2
\end{eqnarray}
for each $Q=P,~V,~S,~A,~D$. These calculations give the parameters
$A_P,~A_V,~A_S,~A_A,~A_D$. On this step quark masses $m_f$ turned out
to be a little bit different for different $Q$.

The second step. The quark masses $m_f$ are determined by
calculation of the minimum
\begin{eqnarray}
\label{Min2}
&& \Omega_{m_q}=\min\limits_{m_q}\sum\limits_{Q}
\sum\limits_{q:u,s,c,b}\left[1-
{M_{qQ}(m_q,m_q,A_Q)\over (M_{qQ})_{\rm exp}}\right]^2
\end{eqnarray}
with parameters $A_Q$ are fixed by (\ref{Min1}).
This calculation give the constituent quark masses
$m_f=m_u,~m_s,~m_c,~m_b$.

The third step. We calculate the parameter $A_T$
\begin{eqnarray}
\label{MinT}
&& \Omega_T=\min\limits_{A_T}\sum\limits_{f,f':u,s}\left[1-
{M_{T,m_f,m_{f'}}(m_f,m_{f'},A_Q)\over
(M_{T,m_f,m_{f'}})_{\rm exp}}\right]^2
\end{eqnarray}
using the constituent quark masses $m_f=m_u,~m_s,~m_c,~m_b$.

The fourth step. We calculate the accuracy of our approximation for
each given quantum number $Q$
\begin{eqnarray}
\label{accur}
&& O_Q={1\over N_Q}\sum\limits_{m_1,m_2:u,s,c,b}\left[1-
{M_{Q,m_1,m_2}(m_1,m_2,A_Q)\over (M_{Q,m_1,m_2})_{\rm exp}}\right]^2
\end{eqnarray}
where $N_Q$ is the number of known mesons with quantum number $Q$.

The results are
\begin{center}
\begin{tabular}{|c|c|c|c|c|c|c|}
\hline
&&&&&&\\
$m_f$ & $m_u$ & $m_s$ & $m_c$ & $m_b$ & &\\
$Mev$& 260 & 434 & 1506 &  4732 & &\\
&&&&&&\\
\hline
&&&&&&\\
$Q$ & $P=0^-$ & $V=1^-$ & $S=0^+$ & $A=1^+$ & $D=2^+$ & $T=3^-$ \\
&&&&&&\\
$A_Q(Gev^2)$& 0.0216 & 0.217 & 0.249 & 0.527 & 0.618 & 0.838 \\
$O_Q$ & 0.011 & 0.0026 & 0.012 & 0.0023 & 0.0019 & 0.00016\\
&&&&&&\\
\hline
\end{tabular}
\end{center}
and are given in the Figures 1-6.

Our results can be formulated in the form:
\begin{itemize}
\item The formula (\ref{MQap}) correctly describes the mass dependence
of mesons on masses of constituent quarks.
\item The main features of meson spectrum are defined by the function
${\cal E}(M,m_1,m_2)$ which does not depend on quantum numbers of a
meson state but on the quark constituent masses.
\item The coupling constant should be small enough $\alpha_s<1$.
\item The mass of a meson should be more then sum of masses of
constituent quarks $M>m_1+m_2$.
\item Masses of constituent quarks $m_q$ are close to the standard
customary values.
\item This consideration is too rough for light pseudoscalar mesons
$\pi,~K,~\eta,~\eta'.$
\item There is large difference between parameters $A_Q$.
\end{itemize}

This paper is supported by RFBR grant $N^\circ~04-0217370$.

\begin{figure}[htb]
\includegraphics{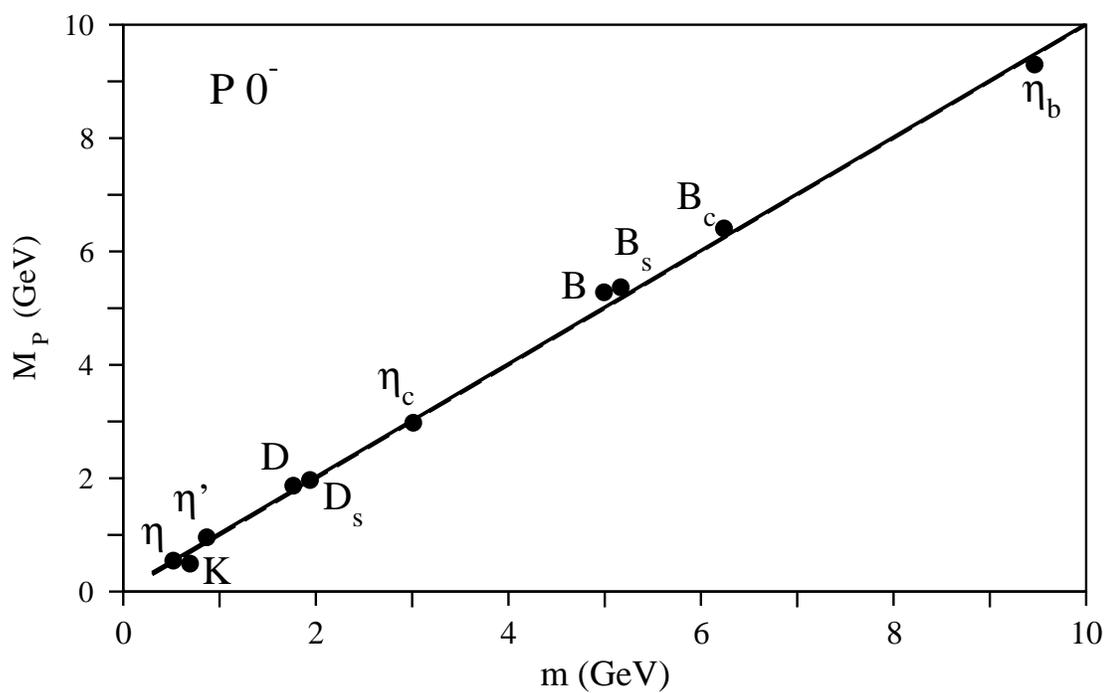}
\caption{Masses of pseudoscalar mesons $P=0^-$. Dotted line is
$M=m_1+m_2=m$, solid line is $M={\cal
M}\left({m\over2},{m\over2},A_P\right)$} 
\label{fig:pseudoscalar}
\end{figure}

\vspace{12mm}

\begin{figure}[htb]
\includegraphics{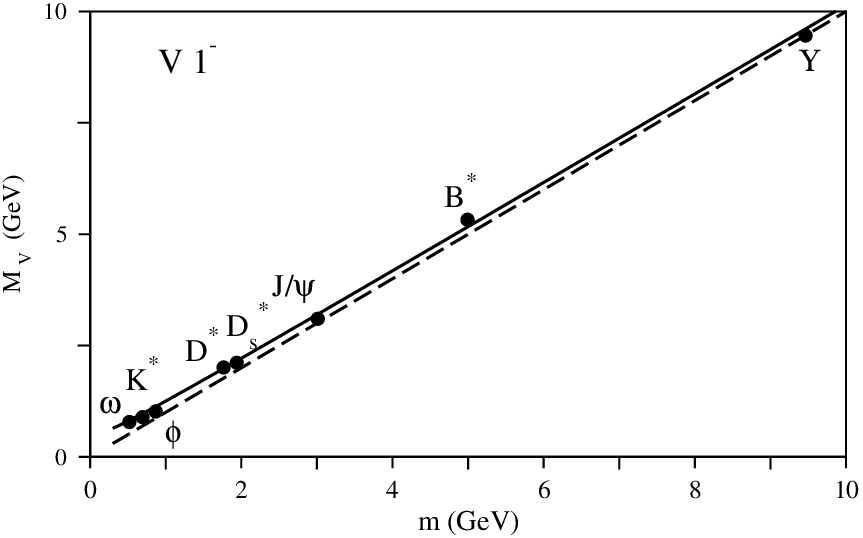}
\caption{Masses of vector mesons $V=1^-$. Dotted line is
$M=m_1+m_2=m$, solid line is $M={\cal M}\left({m\over2},{m\over2},A_V\right)$}
 \label{fig:vector}
\end{figure}

\newpage

\begin{figure}[htb]
\includegraphics{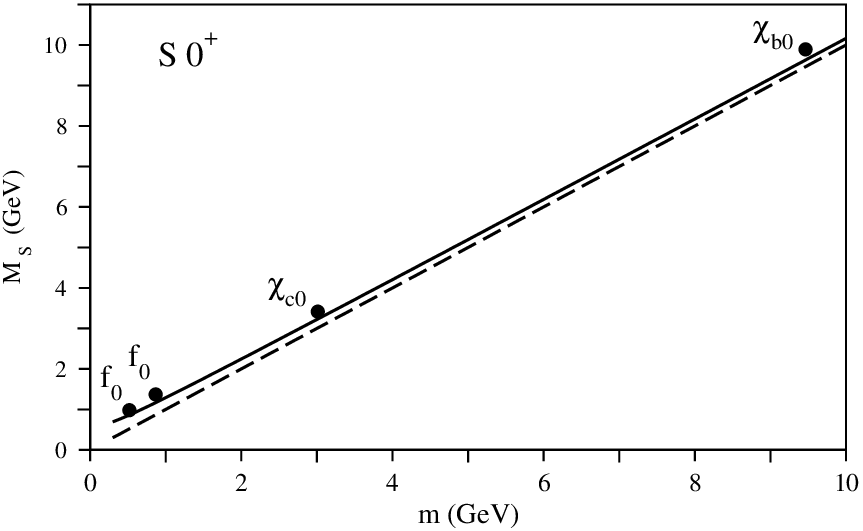}
\caption{Masses of scalar mesons $S=0^+$. Dotted line is
$M=m_1+m_2=m$, solid line is $M={\cal
M}\left({m\over2},{m\over2},A_S\right)$} \label{fig:scalar}
\end{figure}

\vspace{30mm}

\begin{figure}[htb]
\includegraphics{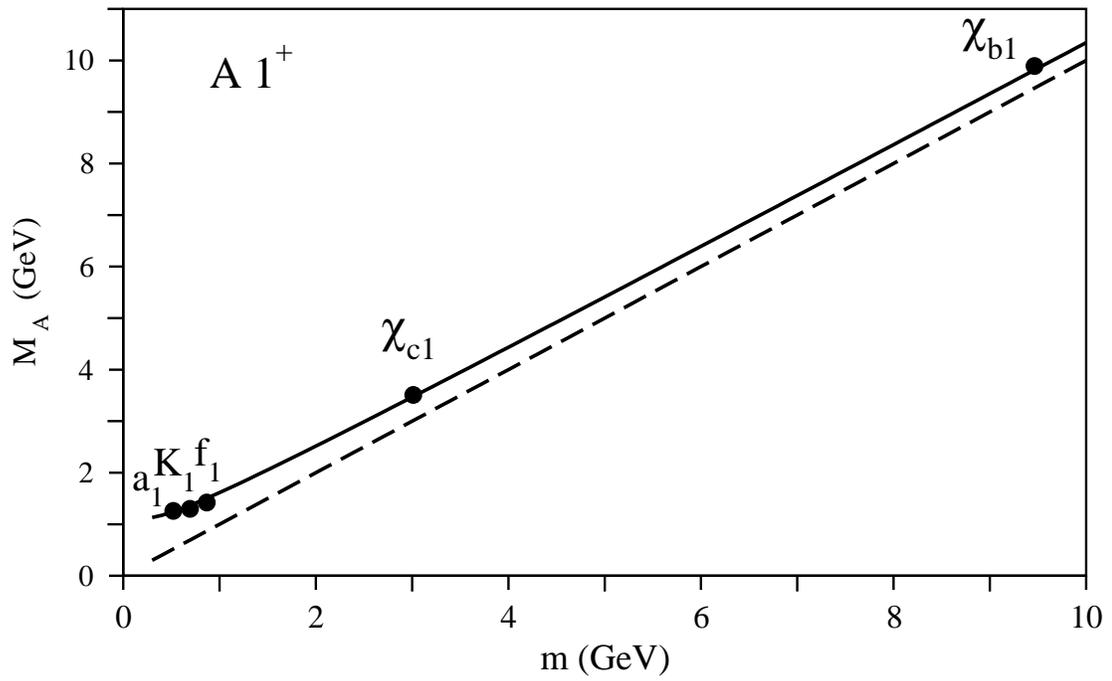}
\caption{Masses of axial vector mesons $A=1^+$. Dotted line is
$M=m_1+m_2=m$, solid line is $M={\cal M}\left({m\over2},{m\over2},A_A\right)$} \label{fig:axial}
\end{figure}

\newpage

\begin{figure}[htb]
\includegraphics{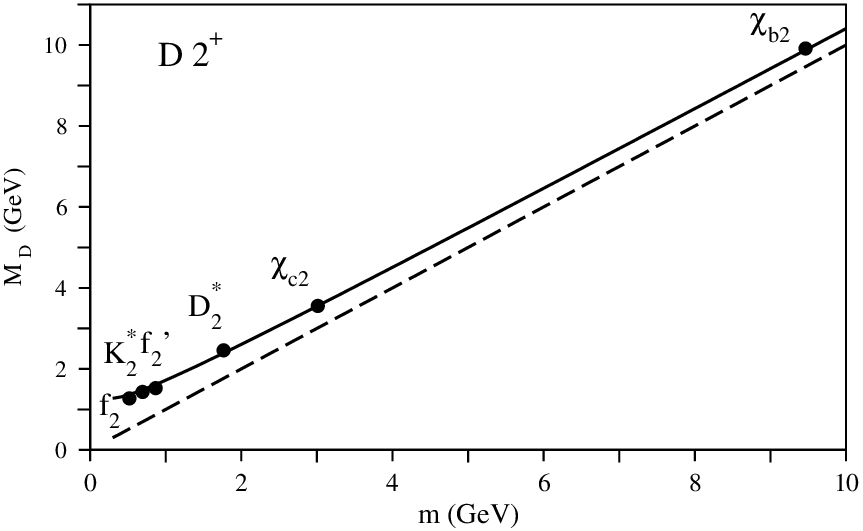}
\caption{Masses of tensor mesons $D=2^+$. Dotted line is
$M=m_1+m_2=m$, solid line is $M={\cal
M}\left({m\over2},{m\over2},A_D\right)$} \label{fig:D}
\end{figure}

\vspace{30mm}

\begin{figure}[htb]
\includegraphics{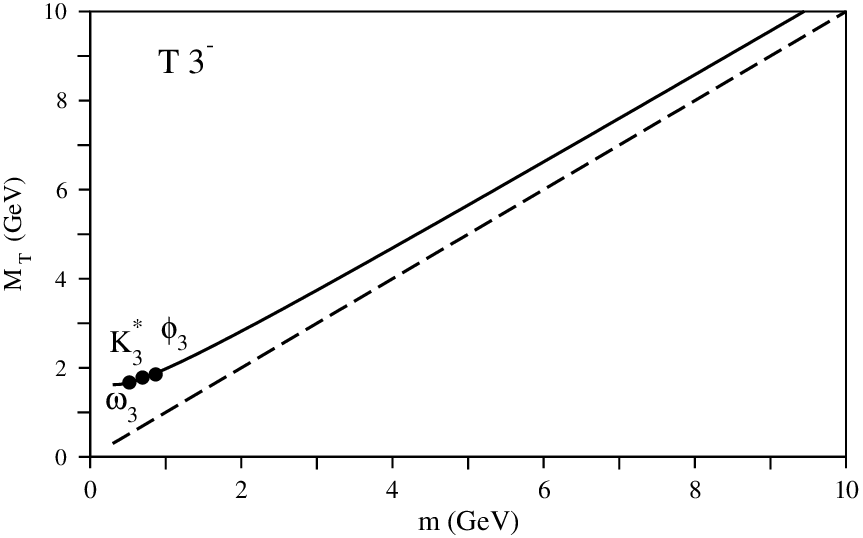}
\caption{Masses of tensor mesons $T=3^-$. Dotted line is
$M=m_1+m_2=m$, solid line is $M={\cal
M}\left({m\over2},{m\over2},A_T\right)$} \label{fig:T}
\end{figure}

\end{document}